# Modeling the History of Astronomy: Ptolemy, Copernicus and Tycho

**Todd Timberlake**
Berry College, Mount Berry, Georgia 30149

**Abstract**

This paper describes a series of activities in which students investigate and use the Ptolemaic, Copernican, and Tychonic models of planetary motion. The activities guide students through using open source software to discover important observational facts, learn the necessary vocabulary, understand the fundamental properties of different theoretical models, and relate the theoretical models to observational data. After completing these activities students can make observations of a fictitious solar system and use those observations to construct models for that system.

## 1. INTRODUCTION

If we are to shape student attitudes about science in an introductory astronomy course, then we must go beyond teaching the results of science and engage students in the scientific process (Wittman 2009, Duncan 2012). Students should have the opportunity to develop and test models of natural phenomena, and even evaluate competing scientific models (Etikina, Warren, & Gentile 2006). One way to involve students in this kind of authentic scientific investigation is to have them investigate scientific models from the history of science (Matthews 1994). This paper describes a series of activities in which students investigate and use models that were important in the historical development of planetary astronomy. The models examined are those introduced by the Hellenistic astronomer Claudius Ptolemy in the 2nd Century, by the Polish astronomer Nicolaus Copernicus in the mid-16th Century, and by the Danish astronomer Tycho Brahe in the late 16th Century. For more information on these theories and the history of planetary astronomy see Kuhn (1985), Crowe (1990), and Linton (2004).

The activities help students use open source software to discover important observational facts, learn the necessary vocabulary, understand the fundamental properties of different theoretical models, and relate the theoretical models to observational data. Once they understand the observations and models, students complete a series of projects in which they observe and model a fictitious solar system with four planets orbiting in circles around a central star. For this purpose, each student is given a different computer program that simulates the motion of the central star and three planets (one of the four planets is the observer's home planet) against a fixed background of stars. Most of the computer simulations were created with *Easy Java Simulations* (Esquembre 2013) and are part of the *Open Source Physics* collection (Open Source Physics 2013). All of the computer simulations, activity handouts, and project materials have been collected in a shared folder in the *Open Source Physics* collection on the *ComPADRE* digital archive (Timberlake 2013).

## 2. OBSERVING THE SKY

The Ptolemaic, Copernican and Tychonic theories were attempts to model naked-eye observations of the night sky. Before exploring these theories, students make simulated observations (both qualitative and quantitative) of the night sky using the open-source planetarium program *Stellarium* (Stellarium 2013).

Observations of the stars lead students to their first astronomical model: the *Celestial Sphere*. The stars appear to move as though they are stuck on the surface of a giant sphere with Earth at the center. The Celestial Sphere rotates east to west once every *sidereal day* (23 hours, 56 minutes) about a fixed axis. For the purpose of examining the motions of all other celestial bodies, the Celestial Sphere can be used as a fixed background. For example, students discover that the Sun drifts roughly eastward relative to the Celestial Sphere along a great circle path known as the *Ecliptic*. As a result of this drift it takes slightly longer (24 hours, or one *solar day*) for the Sun to complete a full rotation in our sky, and the Sun completes one full circuit along the Ecliptic in a sidereal year (about 365.25 solar days).

Students then observe the five visible planets: Mercury, Venus, Mars, Jupiter, and Saturn. They find that all five planets move relative to the Celestial Sphere. The planets move generally eastward and remain near the Ecliptic, although they can be above or below. The average time for a planet to complete one circuit along the Ecliptic is the planet's *zodiacal period*. Planets also move relative to the Sun. The *elongation* of a planet is the angle between the planet and the Sun on the sky. If a planet is at 0° elongation it is said to be in *conjunction*, at 90° it is in *quadrature*, and at 180° it is in *opposition*. The Sun periodically moves eastward past each planet and the time between one pass and the next is that planet's *synodic period*. Students are often surprised to discover that the planets occasionally move westward for a short time, in what is called *retrograde* motion, before resuming their eastward motion. The time between successive retrogrades is observed to be equal to the planet's synodic period.

In spite of these general characteristics, not all planets behave the same. Mercury and Venus are never more than 28° and 48°, respectively, from the Sun. They are in conjunction during the middle of their retrograde motion. These planets are known as *inferior planets*. Mars, Jupiter, and Saturn can attain any elongation. They are in opposition during the middle of their retrograde motion. These planets are known as *superior planets*. All planets appear somewhat brighter during retrograde, but Mars displays the greatest increase in brightness.

After exploring and measuring the real (simulated) night sky, students make observations of the night sky in their personalized solar system. Students must determine the number of planets and classify each planet as inferior or superior. They must measure the sidereal year ($T_{sy}$), as well as the zodiacal period ($T_z$) and synodic period ($T_s$) of each planet. They must measure the maximum elongation ($\alpha$) of each inferior planet and the time from opposition to quadrature ($t_Q$) for each superior planet. These measurements will be used later to construct models of their solar system.

## 3. PTOLEMAIC MODELING

The next set of activities helps students explore a simplified version of Ptolemy's model for planetary motions using the *Inferior Ptolemaic* and *Superior Ptolemaic* EJS programs (Timberlake 2013). In the simplified Ptolemaic model each planet moves uniformly counterclockwise on a circle called the *epicycle* while the center of the epicycle moves uniformly counterclockwise along the *deferent*, a larger circle centered on Earth. The Sun moves uniformly counterclockwise along a circle centered on Earth.[1]

By working with the simulations students discover that the period of the Sun's orbit must equal the sidereal year. They find that planets retrograde when the motion of the planet along the epicycle is in the opposite direction of the motion of the epicycle center along the deferent, which occurs when the planet is on the innermost part of the epicycle. The period of the epicycle center's motion around the deferent must equal the planet's zodiacal period. If the period of the planet's motion around the epicycle is measured relative to the deferent (so that a full period is measured from the time the planet crosses the deferent circle going outward until the next time it does so) then this epicycle period must equal the planet's synodic period.

Students also discover the main difference between the Ptolemaic theories for inferior and superior planets. The center of an inferior planet's epicycle must remain on the Earth-Sun line in order to keep the planet near the Sun in the sky. For superior planets, on the other hand, the line from the center of the epicycle to the planet must always be parallel to the Earth-Sun line in order to ensure that the planet always retrogrades at opposition.

---

[1] In the full Ptolemaic model the Sun's orbit and the deferent circles were not centered on Earth, and the epicycle centers did not move uniformly along the deferent. The theories for Mercury and Venus had additional complications.

Students can also determine the relative sizes of each planet's epicycle and deferent. For an inferior planet, the ratio of its epicycle radius to its deferent radius is given by

$$\frac{R_e}{R_d} = \sin\alpha,  \quad (1)$$

where α is the planet's maximum elongation, as shown in Figure 1.

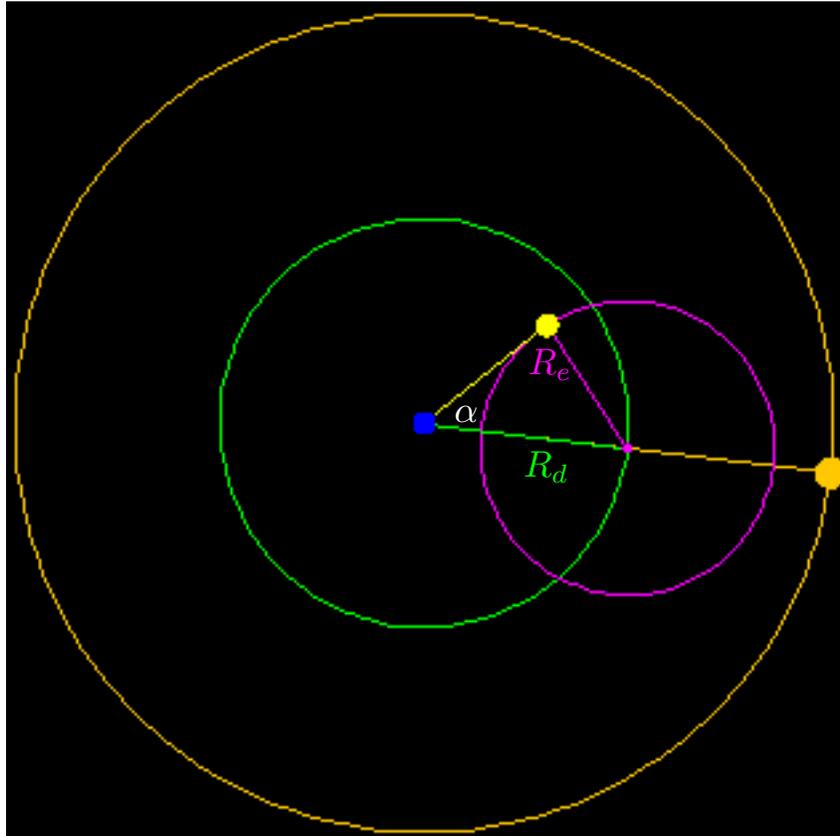

**Figure 1:** Ptolemaic geometry for an inferior planet at maximum elongation. Note that the center of the epicycle lies on the Earth-Sun line.

For a superior planet this ratio can be computed by comparing the geometries at opposition and at quadrature, as shown in Figure 2. The result looks very similar to Equation 1:

$$\frac{R_e}{R_d} = \sin\theta, \quad (2)$$

but note that α has been replaced by θ, an angle that is not observable (because there is nothing visible at the center of the epicycle). By examining the change between opposition and quadrature, and recalling that all motions are uniform, students can show that

$$\theta = 90° + 360°\frac{t_Q}{T_z} - 360°\frac{t_Q}{T_{sy}}, \quad (3)$$

with symbols defined as above.

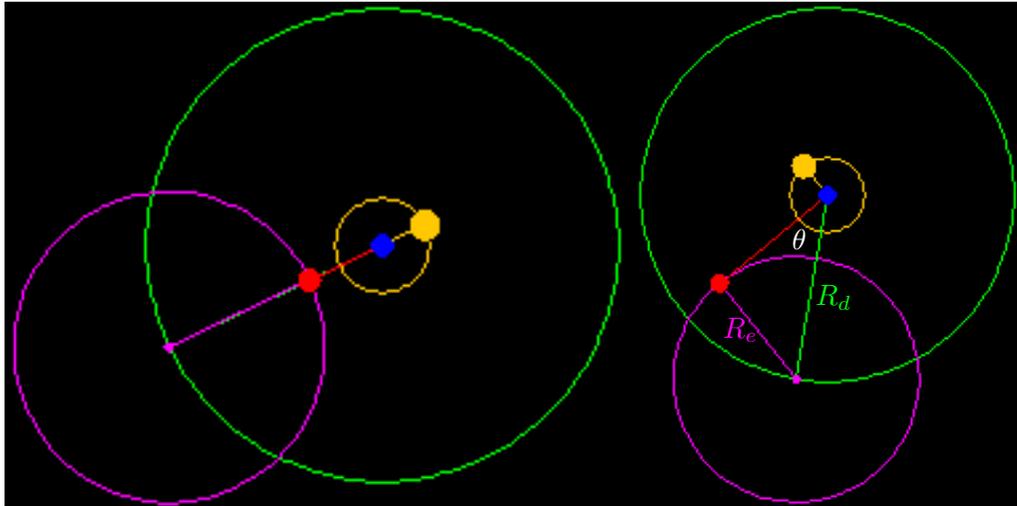

**Figure 2:** Ptolemaic geometry for a superior planet at opposition (left) and at eastern quadrature (right). Note that the line from the center of the epicycle to the planet is always parallel to the Earth-Sun line.

After exploring the simplified Ptolemaic model of our solar system, students can develop a Ptolemaic model for their personalized fictitious solar system using the observational data they collected earlier. Students also are in a position to evaluate Ptolemy's theory. The model matches the observational data both qualitatively and, to some extent, quantitatively. In particular, the model can reproduce retrograde motion and it automatically makes planets brighter during retrograde because they are closer to Earth at that time. However, there are also some odd features of the model. Retrograde can be synchronized to opposition/conjunction only by adding the somewhat mysterious constraints described above. No explanation is offered for why planets come in two different types, with different motions for each type. Although the ratio of a planet's epicycle to its deferent is fixed, there is no set scale for relating the size of one planet's orbit to another. Thus, even the order of the planets is not determined in the Ptolemaic system.

## 4. COPERNICAN MODELING

In the next set of activities students use the *Copernican System* EJS program (Timberlake 2013) to explore a simplified version of the Copernican theory in which each planet moves uniformly counterclockwise on a circle centered on the Sun.[2] Students discover that the Earth, now treated as a planet that rotates to produce the apparent rotation of the Celestial Sphere, must have an orbital period $T_E$ that is equal to one sidereal year. They find that all other planets are naturally classified as inferior or superior depending on whether their orbit is smaller or larger than Earth's, respectively. The simulations help students discover a formula for the period $T_p$ of a planet's orbit:

$$T_p = \left(T_E^{-1} \pm T_s^{-1}\right)^{-1}, \tag{4}$$

where $T_s$ is the planet's synodic period (+ for inferior, - for superior).

---

[2] In the full Copernican system planets moved on small epicycles which in turn moved uniformly on circular orbits that were not centered on the Sun. These small epicycles effectively reproduced the non-uniform motion of the Ptolemaic system. The theories for Mercury and Venus had additional complications.

Students also find that the relative sizes of all planetary orbits can be determined from observations. The ratio of the orbital radius $R_I$ for an inferior planet to the orbital radius $R_E$ of Earth is given by

$$\frac{R_I}{R_E} = \sin\alpha, \tag{5}$$

where α is the planet's maximum elongation as shown in Figure 3.

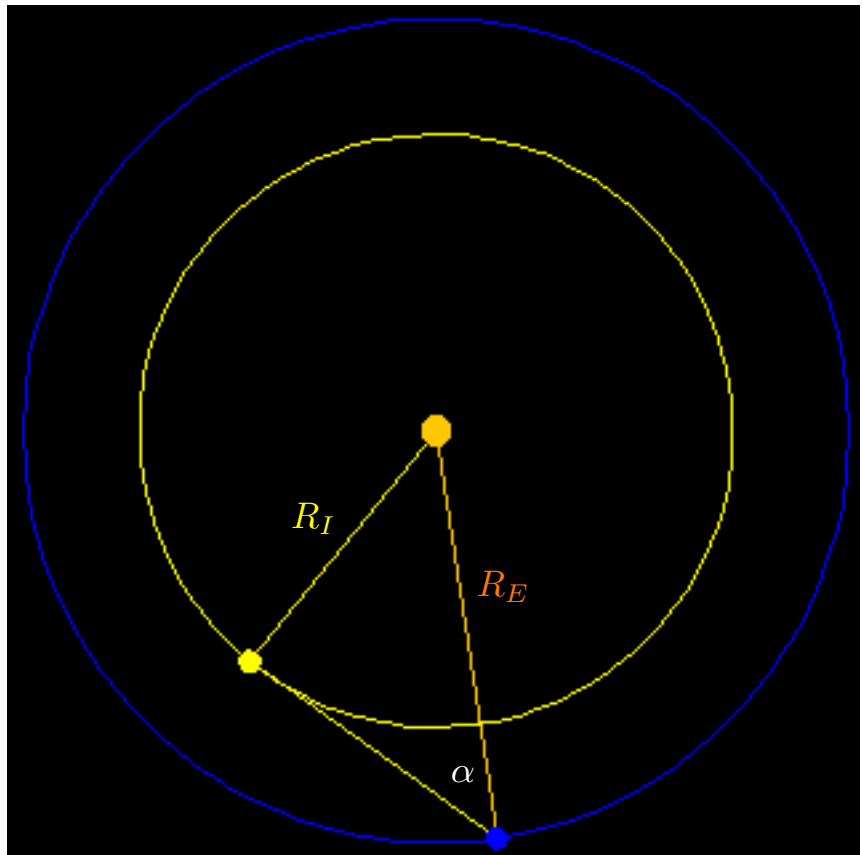

**Figure 3:** Copernican geometry for an inferior planet at maximum elongation.

Once again the formula is more complicated for superior planets. The geometry at quadrature (Figure 4) shows that

$$\frac{R_E}{R_S} = \sin\theta, \tag{6}$$

where $R_S$ is the radius of the superior planet's orbit and $\theta$ is an angle that cannot be measured directly from Earth (it is actually the maximum elongation of Earth as seen from Mars). As in the Ptolemaic system, students can consider the motion from opposition to quadrature to find that

$$\theta = 90° + 360° \frac{t_Q}{T_P} - 360° \frac{t_Q}{T_E} = 90° - 360° \frac{t_Q}{T_S}. \tag{7}$$

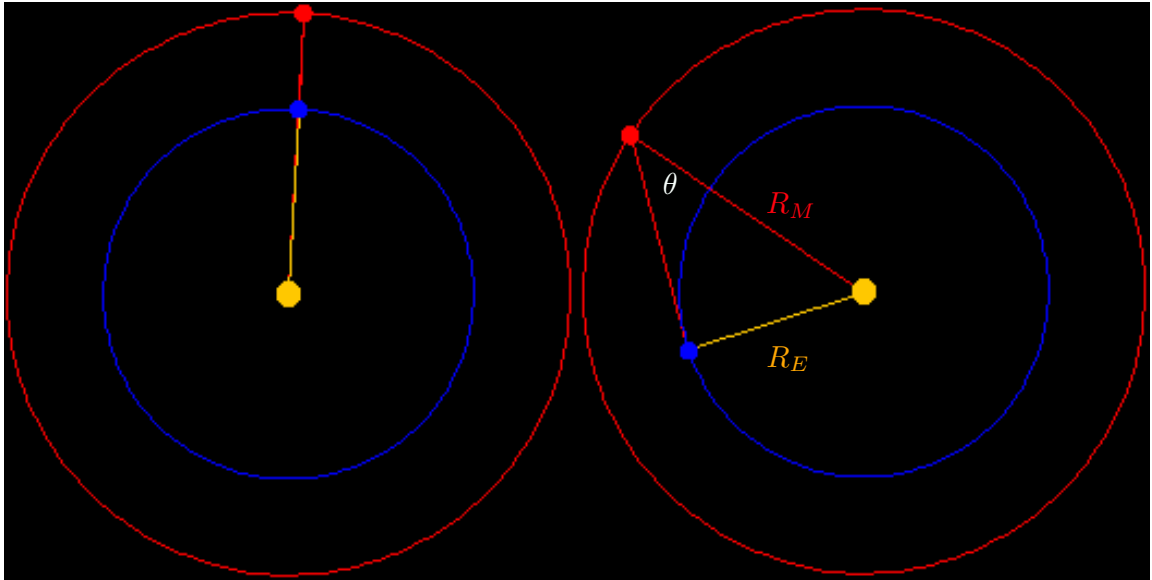

**Figure 4:** Copernican geometry for a superior planet at opposition (left) and eastern quadrature (right).

After exploring the simplified Copernican model of our solar system, students can develop a Copernican model for their personalized fictitious solar system using the observational data they collected earlier. They can also compare the Copernican model to the Ptolemaic model. The Copernican model matches the observational data just as well as the Ptolemaic model, but without any special constraints. The model provides a natural explanation for the distinction between inferior and superior planets. The Copernican model produces retrograde motion in a natural way when Earth passes, or is passed by, another planet. Retrograde motion is automatically synchronized to conjunction/opposition and also automatically occurs when the planet is closest to Earth (and thus brightest). The sizes of all planetary orbits are fixed relative to Earth's orbit, which provides a unique ordering for the planets. The Copernican model also exhibits a harmony between orbital size and orbital speed: planets closer to the Sun orbit at greater speeds and thus have shorter orbital periods.

It might seem that the Copernican model is superior to the Ptolemaic model in many ways, but historically there were serious problems with the Copernican model. It postulated motions of the Earth that were undetectable and which contradicted the established (Aristotelian) physics, as well as common sense. It also predicted an annual parallax of the stars, which was not observed (Timberlake 2012). For these and other reasons the Copernican theory was not readily accepted in spite of its many aesthetically pleasing qualities (Martin 1984).

## 5. PTOLEMY, COPERNICUS, AND TYCHO

The Tychonic theory is essentially a Copernican theory with a stationary Earth. This theory served as a compromise between the Ptolemaic and Copernican systems: it shares many of the aesthetic properties of the Copernican theory, while avoiding the problems of a moving Earth. Students can use the *Ptolemy Copernicus Tycho* EJS program (Timberlake 2013) to demonstrate the geometric equivalence of the simplified versions of these three theories (see Figure 5). Note that the Ptolemaic theory has been scaled so that the deferent of the inferior planet, and the epicycle of the superior planet, are the same size as the Sun's orbit. Ptolemy never would have used this scaling, but it is permitted geometrically.

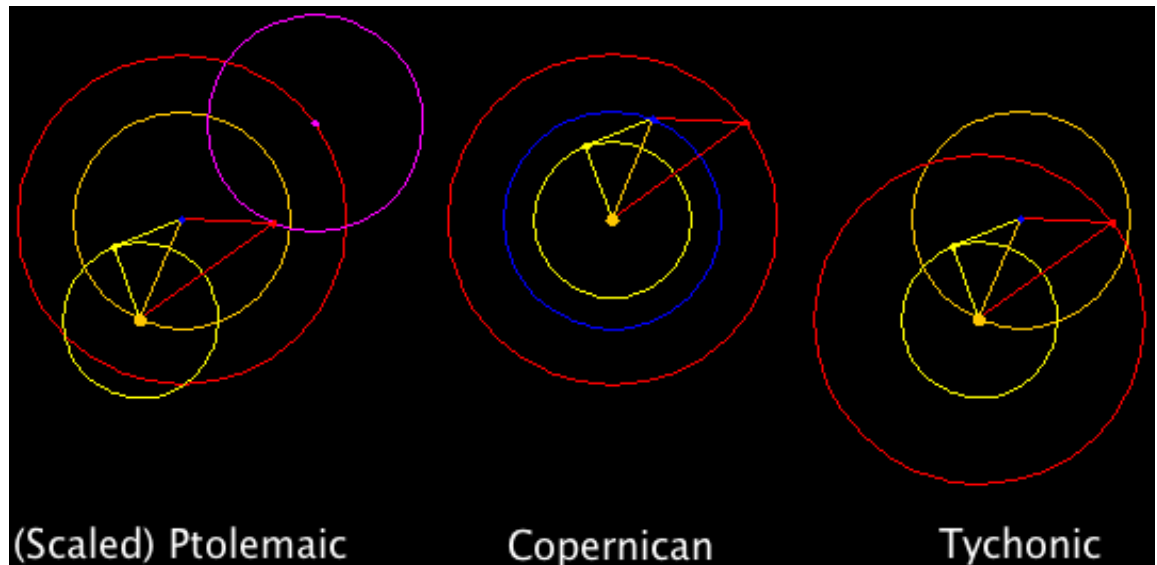

**Figure 5:** Comparison of simplified versions of the (scaled) Ptolemaic, Copernican, and Tychonic systems.

This program also can help students explore the relationships between the theories. For example, the orbit of an inferior planet in the Tychonic model is identical to the same orbit in the Ptolemaic model. For superior planets, the Tychonic model just swaps the epicycle and deferent from the scaled Ptolemaic model. Once they understand these connections between the Tychonic and Ptolemaic models, students can identify the connections between the Copernican and Ptolemaic models (Brehme 1976). These connections are detailed in Table 1, which lists the elements of the Ptolemaic model and the elements of the Copernican model to which they correspond.

Students can verify that these connections are consistent with the results of their earlier modeling. For example, Equations 1 and 5 show that the ratio of an inferior planet's orbital radius to Earth's orbital radius in the Copernican model is equal to the ratio of that planet's epicycle radius to its deferent radius in the Ptolemaic model, as we should expect based on Table 1.

| Ptolemaic Element | Copernican Element |
|---|---|
| Orbit of Sun | Orbit of Earth |
| Inferior Deferent | Orbit of Earth |
| Inferior Epicycle | Orbit of inferior planet |
| Superior Deferent | Orbit of superior planet |
| Superior epicycle | Orbit of Earth |

**Table 1:** Correspondences between the simplified Ptolemaic and Copernican models.

## 6. CONCLUSION

These activities not only allow students to engage in developing and testing scientific models, but they also show students that the same data can lead to very different models that use fundamentally different assumptions. Students gain a deep understanding of the simplified Ptolemaic and Copernican models, including how to construct Ptolemaic and Copernican models using their own observational data. By exploring more than one model of planetary motion, students have the opportunity to evaluate the theories based on empirical adequacy, consistency with other accepted theories, and even aesthetic criteria. In short, students are given the opportunity to engage in the process of doing science.

If these activities and projects are followed by a discussion of how Galileo, Newton and others introduced new ideas about the physics of motion, then students can learn about how developments in one area of science can lead to the re-evaluation of theories in other areas. In the 16$^{th}$ century the Copernican model conflicted with the accepted (Aristotelian) physics and was rejected in favor of the Ptolemaic or Tychonic models. By the 18$^{th}$ century Newtonian physics became the dominant physical theory and the Copernican model became accepted, in spite of the fact that it predicted an as-yet-unobserved annual stellar parallax, because it fit much better with Newtonian ideas about motion than did the Tychonic theory. Studying this episode in the history of astronomy can give students significant insight into how scientific theories are evaluated and how those evaluations change over time, thus helping them gain a better understanding of the nature of science.